\documentclass[preprint]{IEEEtran}
\usepackage{hyperref}
\usepackage{color}
\usepackage{amsmath}
\usepackage{bm}
\usepackage{mathrsfs}
\usepackage{multirow}
\usepackage{subfig}
\usepackage{graphicx}
\usepackage{amssymb}
\usepackage{calc}
\usepackage{stfloats}
\usepackage{comment}
\usepackage{algorithm,algorithmic,textcomp,gensymb,cite}
\usepackage{booktabs}
\usepackage{array}
\usepackage{tabularx}
\usepackage{epstopdf}
\usepackage{multirow}
\usepackage{fancyhdr}
\graphicspath{{Figure/}}

\makeatletter
\def\endthebibliography{%
  \def\@noitemerr{\@latex@warning{Empty `thebibliography' environment}}%
  \endlist
}

\newcounter{MYtempeqncnt1}
\newcounter{MYtempeqncnt2}
\newcounter{MYtempeqncnt3}
\newcounter{MYtempeqncnt4}
\newcounter{MYtempeqncnt5}
\newcounter{MYtempeqncnt6}

\fancypagestyle{firstpage}
{
    \pagestyle{fancy}
    \fancyhead[LO]{\textbf{This work has been submitted to the IEEE for possible publication. Copyright may be transferred without notice, after which this version may no longer be accessible.}}
}

\begin{document}

\title{Variational Nonlinear Kalman Filtering with Unknown Process Noise Covariance}

\author{Hua Lan\thanks{Hua Lan, Jinjie Hu and Zengfu Wang are with the School of Automation, Northwestern Polytechnical University, and the Key Laboratory of Information Fusion Technology, Ministry of Education, Xi'an, Shaanxi, 710072, P. R. China.
Qiang Cheng is with the Nanjing Research Institute of Electronics Technology and the Key Laboratory of IntelliSense Technology, Nanjing, Jiangsu, 210039, P. R. China.
This work was in part supported by the National Natural Science Foundation of China~(Grant No. 61873211 and U21B2008).}, Jinjie Hu, Zengfu Wang*\thanks{*~Corresponding author:~Zengfu Wang.}, Qiang Cheng}
\maketitle
\thispagestyle{firstpage}

\begin{abstract}
Motivated by the maneuvering target tracking with sensors such as radar and sonar, this paper considers the joint and recursive estimation of the dynamic state and the time-varying process noise covariance in nonlinear state space models. Due to the nonlinearity of the models and the non-conjugate prior, the state estimation problem is generally intractable as it involves integrals of general nonlinear functions and unknown process noise covariance, resulting in the posterior probability distribution functions lacking closed-form solutions.
This paper presents a recursive solution for joint nonlinear state estimation and model parameters identification based on the approximate Bayesian inference principle. The stochastic search variational inference is adopted to offer a flexible, accurate, and effective approximation of the posterior distributions.
We make two contributions compared to existing variational inference-based noise adaptive filtering methods.
First, we introduce an auxiliary latent variable to decouple the latent variables of dynamic state and process noise covariance, thereby improving the flexibility of the posterior inference.
Second, we split the variational lower bound optimization into conjugate and non-conjugate parts, whereas the conjugate terms are directly optimized that admit a closed-form solution
and the non-conjugate terms are optimized by natural gradients, achieving the trade-off between inference speed and accuracy. The performance of the proposed method is verified on radar target tracking applications by both simulated and real-world data.
\end{abstract}


\section{Introduction}
In most state estimation problems, it is generally required to perform the nonlinear state estimation in the presence of uncertain model parameters.
For instance, there exists nonlinear coordinate transformation in maneuvering target tracking with remote sensing systems such as radar and sonar~\cite{2003LiSurvey}, whereas the target dynamics are generally modeled in Cartesian coordinates, and sensor measurements are processed in polar coordinates. Meanwhile, the model parameters, especially the process noise covariance, are uncertain, arising from the unexpected maneuver of the non-cooperative target. Due to the nonlinearity of the models and the uncertainty of the model parameters, the state estimation problem is generally intractable as it involves integrals of the general nonlinear functions and the unknown model parameters, resulting in the posterior probability distribution function~(PDF) of the system state lacking closed-form solutions.

The problems of nonlinear state estimation and model parameter identification have their own line of research, which have been active for decades in the target tracking community. Specifically, the nonlinear state estimation techniques can be generally classified into function approximation, moment approximation, and stochastic approximation. Function approximation techniques approximate a stationary non-linear dynamical system with a non-stationary linear dynamical system, such as a first order Taylor series expansion of the extended Kalman filter~\cite{2012GustafssonSome}. Moment approximation techniques directly approximate the posterior PDF by Gaussian approximation, where the moments of Gaussian distribution are computed using numerical integration,
such as unscented transformation of the unscented Kalman filter~\cite{2004JulierUnscented}, spherical cubature integration of the cubature Kalman filter~\cite{arasaratnam2009cubature}, Gauss-Hermite integration of the quadrature Kalman filter~\cite{ito2000gaussian}. Stochastic approximation techniques, such as the particle filter~\cite{Arulampalam2002}, directly approximate the intractable posterior PDF by Monte Carlo integration, i.e., a set of weighted samples called particles. There are three main categories of methods for model identification: input estimator, multiple model estimator, and noise adaptive estimator. The input estimator treats the unknown parameters as extra input variables and estimates them with the system state jointly~\cite{2004MookerjeeReduced}.
Multiple model estimator~\cite{2005RongIMM} performs the joint estimation of model probabilities and system state by using a bank of different hypothetical models representing the different modes of system state.
The noise adaptive estimator assumes that the statistics~(e.g., the mean and covariance) of the noise is nonstationary and unknown, and performs the noise statistics identification and state estimation simultaneously. Traditionally, the linear state estimation with unknown noise statistics is known as adaptive Kalman filtering.

As aforementioned, finding the optimal and analytical solution of nonlinear state estimation in the presence of model parameter uncertainties is intractable.
One should resort to approximate inference.
The sampling-based approximate inference, including sequential Monte Carlo, has been widely used in nonlinear state estimation.
However, it suffers from an expensive computational burden, especially for high-dimensional state estimation. As an approach of approximate inference, variational Bayes~(VB)~\cite{Blei2016variational} yields a deterministic approximation procedure that converts an intractable inference of posterior PDF into optimization.
That is, VB posits a class of tractable variational distributions family over the latent space and tries to find the closest distribution to the posterior PDF in terms of in Kullback-Leibler~(KL) divergence via calculus of variations.
VB-based methods have attracted significant attention in nonlinear state estimation and adaptive filtering applications due to their efficient computation compared to sampling-based methods.

In the aspect of nonlinear state estimation, \v{S}m\'{i}dl and Quinn~\cite{2008SmidlVariational} proposed a VB-based nonlinear state estimation, where VB was used to accelerate the marginalized particle filtering. Frigola~\emph{et al.}~\cite{2015FrigolaVariational} addressed the Bayesian learning of nonparametric nonlinear state-space models, where the nonlinear function was modeled by Gaussian processes with parameters being learned by VB. Gultekin and Paisley~\cite{2017GultekinNonlinear} presented three nonlinear filtering approaches based on three different divergence measures, that is, the forward KL divergence as used in variational inference, the reverse KL divergence as used in expectation-propagation, and the $\alpha-$divergence. Hu~\emph{et al.}~\cite{Yumei2018An} proposed nonlinear filtering based on proximal VB approach~\cite{2015KhanKL}.

In the aspect of noise adaptive filtering, S\"{a}rkk\"{a} and Nummenmaa~\cite{sarkka2009recursive} made the first attempt at VB for joint recursive estimation of dynamic state and the time-varying measurement noise parameters in linear state space models. By choosing the Gaussian and Inverse-Gamma distributions as conjugate prior distributions of latent variables, the joint posterior distribution of the state and the process noise covariance are updated iteratively based on the mean-field VB method. Subsequently, this work was extended to the state estimation~\cite{Huang2017TAC} and smoother~\cite{ardeshiri2015approximate} in the presence of both inaccurate process noise covariance and measurement noise covariance. Different from the work of~\cite{Huang2017TAC} which regarded the covariance of the predicted state as latent variable,
Ma~\emph{et al.}~\cite{2018MaImproved} assumed that the conjugate prior distribution directly depends on the underlying latent variables by inducing auxiliary latent variable, resulting in more accurate state estimation.
S\"{a}rkk\"{a} and Hartikainen~\cite{2015MbalawateAdaptive} extended the work of~\cite{sarkka2009recursive} to nonlinear state space model,
and proposed the adaptive Metropolis algorithm whereas the unknown measurement noise covariance matrix is updated with VB-based adaptive Kalman filter.
By choosing Wishart distribution as the conjugate prior distribution of the unknown information matrix, Dong~\emph{et al.}~\cite{2017DongVariational} proposed a VB adaptive cubature information filter for joint state estimation and measurement noise covariance identification in nonlinear state space models.

However, to our best knowledge, few work has been done on the nonlinear state estimation in the presence of unknown process noise covariance based on the VB approach.
There are two main challenges when using the existing mean-filed VB approach for this problem.
First, it is challenging to design the joint conjugate prior distributions on latent variables of the system state and process noise covariance~\cite{2013SarkkaNonlinear}. Second, due to the nonlinearity, it is intractable to directly maximize the evidence lower bound~(ELBO) by the coordinate ascent algorithm. The non-conjugate prior and nonlinearity make the computation of ELBO intractable. Stochastic optimization~\cite{2013HoffmanStochastic} allows direct optimization of the ELBO by Monte Carlo integration to fit the variational parameters, bridging the gap between VB and Monte Carlo methods~\cite{salimans2015markov, naesseth2018variational}. Paisley \emph{et al.}~\cite{paisley2012variational} proposed a stochastic optimization method for sampling unbiased gradients of ELBO and used control variates to reduce the variance of the stochastic gradient.

%There are two general techniques for improving the flexibility of approximate posteriors~\cite{2019KingmaAn} in the context of natural gradient-based VB, including the auxiliary latent variables~\cite{2015SalimansMarkov} and normalizing flows~\cite{2015RezendeVariational}. By auxiliary latent variables, the auxiliary ELBO is more easily optimized than the traditional ELBO. Stochastic VB~\cite{2013HoffmanStochastic}, a method that uses stochastic gradient ascent methods to fit the variational parameters, helps solve nonlinear state estimation problems.

This paper encapsulates the nonlinear state estimation with uncertain process noise covariance into a variational inference problem. The non-conjugate prior and model nonlinearity disable the existing VB-based noise adaptive filtering approaches.
We introduce an auxiliary latent variable to decouple the system state and process noise covariance, such that the auxiliary ELBO is more easily optimized than the traditional ELBO, which improves the flexibility of the posterior PDF inference.
Meanwhile, we split the ELBO into non-conjugate parts~(arising from the nonlinearity of the system model) and conjugate parts, where the non-conjugate parts are optimized by natural gradients and conjugate parts are optimized by normal gradients, achieving the trade-off between inference speed and accuracy. The performance of the proposed method is verified on radar target tracking applications by both simulated and real data.

The remainder of the paper is organized as follows.
Section \ref{sec:problem} describes the problem formulation of nonlinear state estimation with unknown process noise covariance.
Section \ref{sec:VI} presents the proposed nonlinear adaptive Kalman filtering method based on stochastic search variational inference.
Section \ref{sec:simulation} and Section \ref{sec:real} provides the performance comparison by simulated and real data, respectively.
Finally, Section \ref{sec:conclusion} concludes the paper.

\section{Problem Formation}\label{sec:problem}
Consider the following discrete-time nonlinear state space system
\begin{align}
\bm{x}_k &= \bm{f}_k(\bm{x}_{k-1}) + \bm{v}_{k}, \\
\bm{y}_k &= \bm{h}_k(\bm{x}_k) + \bm{w}_k,
\end{align}
where $\bm{x}_k \in \mathbb{R}^{n_x}$ is the target state, $\bm{y}_k \in \mathbb{R}^{n_y}$ is the sensor measurement; $\bm{f}_k(\cdot)$ is the known nonlinear state transition function, $\bm{h}_k(\cdot)$ is the known nonlinear measurement function; $\bm{v}_k \in \mathbb{R}^{n_x}$ and $\bm{w}_k \in \mathbb{R}^{n_y}$ are independent zero-mean Gaussian process noise vector and measurement noise vector with corresponding covariance matrices $\bm{Q}_k$ and $\bm{R}_k$, respectively. Time is indexed by $k$. The initial state vector $\bm{x}_0$ is assumed to follow Gaussian distribution with mean $\bm{\hat x}_{0|0}$ and covariance matrix $\bm{P}_{0|0}$. Moreover, $\bm{x}_0$, $\bm{v}_k$ and $\bm{w}_k$ are assumed to be mutually uncorrelated.

The model parameters of the above state space system include $\{\bm{Q}_k, \bm{R}_k, \bm{\hat x}_{0|0}, \bm{P}_{0|0}\}$.
The initial state estimate $\bm{\hat x}_{0|0}$ and $\bm{P}_{0|0}$ are often less important since they will get ``washed away'' by the data after a few time steps.
The nonlinear measurement function $\bm{h}_k(\cdot)$ is sensor-dependent and typically known in advance.
We assume $R_k$ is known in this paper. 
For maneuvering target tracking, there exists target motion uncertainty.
A tracker does not have access to an accurate dynamic model of the target being tracked, making it difficult to predict the target's motion accurately.
There are mainly three maneuvering motion models: equivalent-noise model, multiple models, and unknown-input model. This paper focuses on the equivalent-noise model, whereas the nonlinear state transition function $\bm{f}_k(\cdot)$ is known for describing typical target trajectories and adapt the unknown and time-varying parameter $\bm{Q}_k$ to describe the target maneuver.

The goal of joint state estimation and model parameter identification is to compute the joint posterior PDF $p(\bm{x}_k, \bm{Q}_k|\bm{y}_{1:k})$.
Formally, the well-known recursive Bayesian filtering solution consists of the following predict-update cycle:
\begin{itemize}
  \item \emph{Initialization}: The recursion starts from the prior distribution $p(\bm{x}_0, \bm{Q}_0)$.
  \item \emph{Prediction}: The one-step-ahead predicted PDF for the joint latent variables $p(\bm{x}_k, \bm{Q}_k|\bm{y}_{1:k-1})$ is given by the Chapman-Kolmogorov equation:
        \begin{equation}\label{eq::prediction}
            \begin{split}
                p(\bm{x}_k, \bm{Q}_k|\bm{y}_{1:k-1}) \!\! = \!\!\!\!\int \!\!&p(\bm{x}_k|\bm{x}_{k-1}, \bm{Q}_k)p(\bm{Q}_k|\bm{Q}_{k-1}) \cdot \\ 
                & p(\bm{x}_{k-1}, \bm{Q}_{k-1}|\bm{y}_{1:k-1})\text{d}\bm{x}_{k-1}\text{d}\bm{Q}_{k-1}.
            \end{split}
        \end{equation}
  \item \emph{Update}: The above predicted PDF is updated by measurement $\bm{y}_k$ using the Bayes rule:
            \begin{equation}\label{eq::update}
                p(\bm{x}_k, \bm{Q}_k|\bm{y}_k) \propto p(\bm{y}_k|\bm{x}_k, \bm{R}_k)p(\bm{x}_k, \bm{Q}_k|\bm{y}_{1:k-1}).
            \end{equation}
\end{itemize}

In the case of linear Gaussian with known model parameters, the posterior PDF of state $\bm{x}_k$ has a closed-form analytical solution via Kalman filtering. For nonlinear functions $\bm{f}_k(\cdot), \bm{h}_k(\cdot)$ and unknown process noise covariance $\bm{Q}_k$,
all the possible values of $\bm{Q}_k$ have to be considered in the prediction step,
and the update step using measurements has no closed-form solution due to the intractable integral computation arising from the nonlinearity in $\bm{f}_k(\cdot), \bm{h}_k(\cdot)$.
One must resort to approximation to perform Bayesian inference in practice. Two main techniques can be used to approximate the intractable joint posterior distribution of $\bm{x}_k$ and $\bm{Q}_k$, including SMC-based sampling approximation and VB-based deterministic approximation.
The former tends to be more accurate but has higher computational costs.
The latter tends to be faster, especially in the case of high-dimensional latent variables. In the next section, we solve the above recursion steps using the VB inference method.

\section{Nonlinear Noise Adaptive Kalman Filtering}\label{sec:VI}
\subsection{Variational Bayesian Inference}
As aforementioned, the difficulty of the nonlinear noise adaptive Kalman filtering arises from solving the intractable joint posterior PDF $p(\bm{x}_k, \bm{Q}_k|\bm{y}_{1:k})$, which consists of the recursive steps in Eqs.~(\ref{eq::prediction})-(\ref{eq::update}). This intractable inference can be approximated by VB methods. Here, we briefly introduce VB and stochastic search VB~(SSVB)~\cite{paisley2012variational} for the intractable ELBO optimization.

Let $p(\bm{z}_{k}, \bm{y}_{k})$ be the probabilistic models for joint latent variables $\bm{z}_{k}$~(e.g., $\bm{x}_{k}$ and $\bm{Q}_{k}$) and observed variables $\bm{y}_{k}$. In Bayesian inference, the key is to compute the posterior PDF $p(\bm{z}_{k}|\bm{y}_{k})$ recursively.
VB transforms the posterior inference into an optimization problem~\cite{Blei2016variational}. That is, VB posits a class of distributions $q(\bm{z}_{k}; \bm{\lambda}_k)$ over latent variables with variational parameters $\bm{\lambda}_k$, and tries to find the closest distribution to the exact posterior distributions $p(\bm{z}_{k}|\bm{y}_{k})$. Closeness is measured by the Kullback-Leibler~(KL) divergence, and the closest distribution is called \emph{variational distribution}. Minimizing the KL divergence from the variational distribution $q(\bm{z}_{k}; \bm{\lambda}_k)$ to the posterior distribution $p(\bm{z}_{k}|\bm{y}_{k})$ is equivalent to maximizing the evidence lower bound~(ELBO) $\mathcal{B}(\bm{\lambda}_k)$. Thus, the objective of VB is to solve the following optimization problem,
\begin{equation}\label{eq::ELBO}
\begin{split}
    p(\bm{z}_{k}|\bm{y}_{k}) & \approx q(\bm{z}_{k}; \bm{\lambda}_k^*) \\
    &= \arg \max_{\bm{\lambda}_k}  \underbrace{\mathbb{E}_{q(\bm{z}_{k}; \bm{\lambda}_k)}\left[\log p(\bm{z}_{k}, \bm{y}_{k}) - \log q(\bm{z}_{k}; \bm{\lambda}_k) \right]}_{\mathcal{B}(\bm{\lambda}_k)}.
\end{split}
\end{equation}

The choice of variational family $q(\bm{z}_{k}; \bm{\lambda}_k)$ trades off the fidelity of the posterior approximation with the complexity of variational optimization. The simplest~(classic) variational family of distributions is the \emph{mean-field} variational family where the latent variables are mutually independent and governed by their parameters $\bm{\lambda}_{i,k}$. For conjugate exponential models, the mean-field VB often has the closed-form solution of Eq.~(\ref{eq::ELBO}) by the coordinate ascent, which repeatedly cycles through and optimizes with respect to each variational parameter $\bm{\lambda}_{i,k}$.
%Though it enables efficient inference, the mean-field family cannot capture posterior dependencies between latent variables and thus decreases the fidelity of the approximation. 
%Moreover, 
Due to the nonlinearity existing in the joint distribution $p(\bm{z}_{k}, \bm{y}_{k})$, the optimization of variational objectives is sometimes intractable since the expectations $\mathbb{E}_{q(\bm{z}_{k}; \bm{\lambda}_k)}\log p(\bm{z}_{k}, \bm{y}_{k})$ do not have a closed-form solution.

SSVB directly optimizes the intractable variational objective function $\mathcal{B}(\bm{\lambda}_k)$ based on stochastic optimization. Assume the ELBO $\mathcal{B}(\bm{\lambda}_k)$ can be separated into two parts, $\mathbb{E}[f(\bm{z}_k)]$ and $h(\bm{z}_k, \bm{\lambda}_k)$, where $f(\bm{z}_k)$ is the function of $\bm{z}_k$ whose expectation is intractable, and $h(\bm{z}_k, \bm{\lambda}_k)$ is the rest term in $\mathcal{B}(\bm{\lambda}_k)$ whose expectation is tractable with close-form solution. The first step in solving the Eq.~(\ref{eq::ELBO}) is to take the gradient of the ELBO $\mathcal{B}(\bm{\lambda}_k)$,
\begin{equation}\label{eq::gradientB}
  \nabla \mathcal{B}(\bm{\lambda}_k) = \nabla_{\bm{\lambda}_k}\mathbb{E}_{q(\bm{z}_{k}; \bm{\lambda}_k)}[f(\bm{z}_{k})] + \nabla_{\bm{\lambda}_k}h(\bm{z}_k, \bm{\lambda}_k).
\end{equation}
The goal of SSVB is to make a stochastic approximation of the first gradient $\nabla_{\bm{\lambda}_k}\mathbb{E}_{q(\bm{z}_{k}; \bm{\lambda}_k)}[f(\bm{z}_{k})]$, while the second gradient is tractable.

Use the identity $\nabla q(\bm{z}_{k}; \bm{\lambda}_k) = q(\bm{z}_{k}; \bm{\lambda}_k) \nabla \log q(\bm{z}_{k}; \bm{\lambda}_k)$. The gradient of the expectation can be rewritten as an expectation with respect to the variational distribution, i.e.,
\begin{equation}\label{eq::gradientC}
  \nabla_{\bm{\lambda}_k}\mathbb{E}_{q(\bm{z}_{k}; \bm{\lambda}_k)}[f(\bm{z}_{k})] = \mathbb{E}_{q(\bm{z}_{k}; \bm{\lambda}_k)}[f(\bm{z}_{k}) \nabla_{\bm{\lambda}_k} \log q(\bm{z}_{k}; \bm{\lambda}_k)].
\end{equation}
With this equation in hand, we can stochastically approximate this expectation using Monte Carlo integration with samples extracted from the variational distribution,
\begin{equation}\label{eq::gradientD}
  \nabla_{\bm{\lambda}_k}\mathbb{E}_{q(\bm{z}_{k}; \bm{\lambda}_k)}[f(\bm{z}_{k})] = \dfrac{1}{N} \sum_{i=1}^N f(\bm{z}^i_{k}) \nabla_{\bm{\lambda}_k} \log q(\bm{z}^i_{k}; \bm{\lambda}_k),
\end{equation}
where $\bm{z}^i_{k} \overset{iid}{\sim} q(\bm{z}_{k}; \bm{\lambda}_k)$, $i = 1, \ldots, N$, and $N$ is the number of samples.

Thus, the gradients of ELBO $\nabla \mathcal{B}(\bm{\lambda}_k)$ is now tractable by replacing the intractable term $\nabla_{\bm{\lambda}_k}\mathbb{E}_{q(\bm{z}_{k}; \bm{\lambda}_k)}[f(\bm{z}_{k})]$ with the unbiased stochastic approximation in Eq.~(\ref{eq::gradientD}). Finally, stochastic optimization updates the variational parameters $\bm{\lambda}_k$ at the $n$-th iteration with
\begin{equation}\label{eq::lambda}
  \bm{\lambda}_k^{(n+1)} = \bm{\lambda}_k^{(n)} + \bm{\rho}_k^{(n)} \nabla  \mathcal{B}(\bm{\lambda}_k^{(n)}),
\end{equation}
which can be carried out by the Adam algorithm~\cite{kingma2015adam},
where the learning rate $\bm{\rho}_k^{(n)}$ follows the Robbins-Monro conditions that $\sum_{n = 1}^{\infty} \bm{\rho}_k^{(n)} = \infty$ and $\sum_{n=1}^{\infty} (\bm{\rho}_k^{(n)})^2 < \infty$. By decreasing the learning rate $\bm{\rho}_k^{(n)}$, the ELBO $\mathcal{B}(\bm{\lambda}_k)$  converges to a local optimal solution.

The practical issue of the stochastic approximation to optimize the ELBO is that the variance of the gradient approximation~(under the Monte Carlo estimation in Eq.~(\ref{eq::gradientD})) can be too large to be useful. In order to decrease the variance, large $N$ and small learning rate $\bm{\rho}_k$ are needed, leading to slow convergence. Variance reduction methods work by replacing the function whose expectation is being approximated via Monte Carlo with another function that has the same expectation but a smaller variance.
Toward this end, one can introduce a control variate $g(\bm{z}_k)$ which approximates $f(\bm{z}_k)$ well but has a closed-form expectation under $q(\bm{z}_k)$~\cite{paisley2012variational}.
Using $g(\bm{z}_k)$ and a scalar $\alpha_k \in \mathbb{R}$, the new function $\hat f(\bm{z}_k)$,
\begin{equation}\label{eq::newFunction}
  \hat f(\bm{z}_k) = f(\bm{z}_k) - \alpha_k\left(g(\bm{z}_k) - \mathbb{E}_{q(\bm{z}_k)}[g(\bm{z}_k)]\right),
\end{equation}
has the same expectations as $f(\bm{z}_k)$ but has a smaller variance. The scalar $\alpha_k = \text{Cov}(f, g) / \text{Var}(g)$ is set to minimize the variance. Therefore, one can replace $f(\bm{z}_k)$ with $\hat f(\bm{z}_k)$ in Eq.~(\ref{eq::gradientD}).

\subsection{Prior Distribution and Auxiliary Latent Variable}
Recall that the recursive  Bayesian filtering consists of  initialization, recursive steps of state prediction and update.
The initialization step is to model the prior distribution of latent variables. In our models, the prior distribution of the system state $\bm{x}_k$, i.e., predicted PDF, is assumed to be Gaussian distribution, i.e.,
\begin{equation}\label{Prior-X}
  p(\bm{x}_k|\bm{y}_{1:k-1}, \bm{Q}_k) = \mathcal{N}(\bm{x}_k| \bm{\hat x}_{k|k-1}, \bm{P}_{k|k-1}),
\end{equation}
where $\bm{\hat x}_{k|k-1}$ and $\bm{P}_{k|k-1}$ are the predicted state and corresponding covariance at time $k$.

The prior distribution of process noise covariance $\bm{Q}_k$ is assumed to be inverse Wishart distribution~\cite{2017DongVariational,ardeshiri2015approximate}, i.e.,
\begin{equation}\label{Prior-Q}
  p(\bm{Q}_k|\bm{y}_{1:k-1}) = \text{IW}(\bm{Q}_k|\hat u_{k|k-1}, \bm{U}_{k|k-1}),
\end{equation}
where $\text{IW}(\bm{P}|\nu, \bm{A})$ signifies that $\bm{P}$ follows an inverse Wishart distribution with degrees of freedom parameter $\nu$ and positive-definite scale matrix $\bm{A} \in \mathbb{R}^{p \times p}$~(with $\nu > p + 1$). The mean value of $\bm{P}$ is $\mathbb{E}[\bm{P}] = \bm{A} / (\nu - p - 1)$. Meanwhile, the inverse $\bm{P}^{-1}$ follows Wishart distribution and has the mean $\mathbb{E}[\bm{P}^{-1}] = \nu \bm{A}^{-1}$.
The initial process noise covariance $\bm{Q}_0$ is assumed to follow an inverse Wishart distribution $\text{IW}(\bm{Q}_0|\hat u_{0|0}, \bm{U}_{0|0})$ with mean value $\bm{\bar Q}_0$ given by
\begin{equation}
  \bm{\bar Q}_0 = \dfrac{\bm{U}_{0|0}}{{\hat u}_{0|0} - n_x - 1},
\end{equation}
where the initial value ${\hat u}_{0|0} = \tau + n_x + 1$ and $\bm{U}_{0|0} = \tau \bm{\bar Q}_0$, where $\tau$ is the tuning parameter.

Combining Eq.~(\ref{Prior-X}) with Eq.~(\ref{Prior-Q}), the joint prior distribution of $\bm{x}_k$ and $\bm{Q}_k$ is given by
\begin{equation} \label{eq::jointPrior}
\begin{split}
& p(\bm{x}_k, \bm{Q}_k|\bm{y}_{1:k-1}) \\
& = \mathcal{N}(\bm{x}_k|\bm{\hat x}_{k|k-1}, \bm{P}_{k|k-1}) \times \text{IW}(\bm{Q}_k|\hat u_{k|k-1}, \bm{U}_{k|k-1}).
\end{split}
\end{equation}

From Eq.~(\ref{eq::jointPrior}), it is seen that the latent variables $\bm{x}_k$ and $\bm{Q}_k$ are dependent since the covariance of predicted state $\bm{P}_{k|k-1}$ is related to $\bm{Q}_k$. Moreover, the joint prior distribution $p(\bm{x}_k, \bm{Q}_k|\bm{y}_{1:k-1})$ is non-conjugate. The work of \cite{sarkka2009recursive, 2015MbalawateAdaptive,2017DongVariational} using the mean-field VB to joint estimation of state $\bm{x}_k$ and measurement noise covariance $\bm{R}_k$ is inappropriate to model the maneuvering target tracking problem. 
In order to obtain the conjugate prior on joint $\bm{x}_k$ and $\bm{P}_{k|k-1}$, the work of \cite{Huang2017TAC} regarded $\bm{P}_{k|k-1}$ as a latent variable, and updated the joint posterior PDFs using VB,
which is tractable but indirect since the unknown parameters are $\bm{Q}_k$ rather than $\bm{P}_{k|k-1}$.

One way to improve the flexibility of the VB models is by introducing \emph{auxiliary latent variables}, which can be used to reduce correlation between the original variables~\cite{lew2022recursive}. In order to decompose $\bm{P}_{k|k-1}$ in Eq.~(\ref{eq::jointPrior}), the continuous auxiliary latent variable $\bm{m}_k$ is introduced.
Then, Eq.~(\ref{Prior-X}) can be rewritten as
\begin{equation}\label{Prior-m}
\begin{split}
   & p(\bm{x}_k|\bm{y}_{1:k-1}, \bm{Q}_k) \\
   &= \int p(\bm{x}_k, \bm{m}_k|\bm{y}_{1:k-1}, \bm{Q}_k) \text{d}\bm{m}_k \\
   &= \int p(\bm{x}_k|\bm{m}_k, \bm{Q}_k) p(\bm{m}_k|\bm{y}_{1:k-1}) \text{d}\bm{m}_k \\
   &= \int \mathcal{N}(\bm{x}_k|\bm{m}_k, \bm{Q}_k) \mathcal{N}(\bm{m}_k|\bm{\hat m}_{k|k-1}, \bm{\Sigma}_{k|k-1})\text{d}\bm{m}_k
\end{split}
\end{equation}
with
\begin{equation}\label{eq::predM}
\begin{split}
  &\bm{\hat m}_{k|k-1} = \mathbb{E}_{p(\bm{x}_{k-1})}[f(\bm{x}_{k-1})], \\
  &\bm{\Sigma}_{k|k-1} \\
  &= \mathbb{E}_{p(\bm{x}_{k-1})}[(f(\bm{x}_{k-1}) - \bm{\hat m}_{k|k-1})(f(\bm{x}_{k-1}) - \bm{\hat m}_{k|k-1})^T].
\end{split}
\end{equation}

The auxiliary latent variable $\bm{m}_k \sim \mathcal{N}(\bm{m}_k|\bm{\hat m}_{k|k}, \bm{\Sigma}_{k|k})$ splits $\bm{P}_{k|k-1}$ into two parts.
Now the process noise covariance $\bm{Q}_k$ is the covariance of the system state $\bm{x}_k$, making the conjugate prior distribution with Gaussian and Inverse Wishart possible.

The objective of VB-based method with auxiliary latent variables $\bm{m}_k$ is to approximate the intractable joint posterior distribution $p(\bm{x}_k, \bm{Q}_k, \bm{m}_k|\bm{y}_k)$ with the tractable variational distribution $q(\bm{x}_k, \bm{Q}_k, \bm{m}_k; \bm{\lambda}_k)$.
The KL-divergence with auxiliary latent variables can be expressed as
\begin{equation}\label{eq:NewELBO}
\begin{split}
&D_{\text{KL}}(q(\bm{x}_k, \bm{Q}_k, \bm{m}_k)||p(\bm{x}_k, \bm{Q}_k, \bm{m}_k)) \\
= & D_{\text{KL}}(q(\bm{x}_k, \bm{Q}_k)||p(\bm{x}_k, \bm{Q}_k)) \\
& + \mathbb{E}_q(D_{\text{KL}}(q(\bm{m}_k|\bm{x}_k, \bm{Q}_k)||q(\bm{m}_k|\bm{x}_k, \bm{Q}_k)).
\end{split}
\end{equation}
From Eq.~\eqref{eq:NewELBO}, it is seen that the ELBO becomes less tight by augmenting an auxiliary latent variable.
However, the auxiliary latent variables can lead to significant improvements on approximating the true posterior PDF since one can access to much more flexible class of variational distribution~\cite{salimans2015markov, ranganath2016hierarchical, maaloe2016auxiliary, lew2022recursive}.

The conjugate prior distributions of joint latent variables $\bm{x}_k$, $\bm{Q}_k$ and $\bm{m}_k$ are given by
\begin{equation}
\begin{split}
& p(\bm{x}_k, \bm{Q}_k, \bm{m}_k|\bm{y}_{1:k-1}) \\
= & \mathcal{N}(\bm{x}_k|\bm{m}_k, \bm{Q}_{k}) \text{IW}(\bm{Q}_k|\hat u_{k|k-1}, \bm{U}_{k|k-1}) \\
  &\times \mathcal{N}(\bm{m}_k|\bm{\hat m}_{k|k-1}, \bm{\Sigma}_{k|k-1}).
\end{split}
\end{equation}
Next, we will determine the prior parameters $\bm{\hat m}_{k|k-1}$, $\bm{\Sigma}_{k|k-1}$ and $\hat u_{k|k-1}$, $\bm{U}_{k|k-1}$.

According to the definition by Eq.~(\ref{eq::predM}) and using the Monte Carlo integration, we have
\begin{equation}\label{eq::predMFinal}
\begin{split}
  \bm{\hat m}_{k|k-1} =& \dfrac{1}{N}\sum_{i=1}^Nf(\bm{x}^i_{k-1}), \\
  \bm{\Sigma}_{k|k-1} =& \dfrac{1}{N}\sum_{i=1}^N(f(\bm{x}^i_{k-1}) - \bm{\hat m}_{k|k-1})(f(\bm{x}^i_{k-1}) - \bm{\hat m}_{k|k-1})^T
\end{split}
\end{equation}
with $\bm{x}_{k-1}^i \overset{iid}{\sim} q(\bm{x}_{k-1}), i = 1, \ldots, N$ and $N$ being the number of samples.

In the vein of \cite{Huang2017TAC, sarkka2009recursive}, the a prior parameters $\hat u_{k|k-1}$, $\bm{U}_{k|k-1}$ are given by
\begin{equation}\label{eq::predQFinal}
\begin{split}
  {\hat u}_{k|k-1} =&  \beta({\hat u}_{k-1|k-1} - n_x - 1) + n_x + 1,\\
  \bm{U}_{k|k-1} =& \beta\bm{U}_{k-1|k-1},
\end{split}
\end{equation}
where $\beta \in (0, 1]$ is the factor for spreading.

\subsection{Approximated Posterior Distribution Update}
We use the mean-field variational family where the latent variables are mutually independent, each governed by a distinct factor in the variational density. Thus, the variational distribution $q(\bm{x}_k, \bm{Q}_k, \bm{m}_k; \bm{\lambda}_k)$ approximates $p(\bm{x}_k, \bm{Q}_k, \bm{m}_k|\bm{y}_{1:k})$ with a free form factorization, i.e.,
\begin{equation}
\begin{split}
& q(\bm{x}_k, \bm{Q}_k, \bm{m}_k;\bm{\lambda_k}) \\
=& q(\bm{x}_k; \bm{\hat x}_{k|k}, \bm{P}_{k|k}) q(\bm{Q}_k; {{\hat u}_{k|k}, \bm{U}_{k|k}}) q(\bm{m}_k; \bm{\hat m}_{k|k}, \bm{\Sigma}_{k|k})
\end{split}
\end{equation}
with $\bm{\lambda}_k = \{\bm{\hat x}_{k|k}, \bm{P}_{k|k}, {\hat u}_{k|k}, \bm{U}_{k|k}, \bm{\hat m}_{k|k}, \bm{\Sigma}_{k|k} \}$.

The optimal variational parameters $\bm{\lambda}_k^*$ can be obtained by maximizing the ELBO $\mathcal{B}(\bm{\lambda}_k)$, i.e.,
\begin{equation}\label{Optimal-Lambda}
\bm{\lambda}_k^* = \arg \max_{\bm{\lambda}_k} \mathcal{B}(\bm{\lambda}_k),
\end{equation}
\begin{equation}
\begin{split}
\nonumber
\mathcal{B}(\bm{\lambda}_k) = \int & q(\bm{x}_k, \bm{Q}_k, \bm{m}_k;\bm{{\lambda}_k}) \cdot \\
& \log \dfrac{p(\bm{x}_k, \bm{Q}_k, \bm{m}_k, \bm{y}_k|\bm{y}_{1:k-1})}{q(\bm{x}_k, \bm{Q}_k, \bm{m}_k;\bm{\lambda_k})} \text{d}\bm{x}_k \text{d}\bm{Q}_k \text{d}\bm{m}_k
\end{split}
\end{equation}
%\begin{equation}\label{Optimal-Lambda}
%\begin{split}
%  \bm{\lambda}_k^* = \arg \max_{\bm{\lambda}_k} \underbrace{ \int q(\bm{x}_k, \bm{Q}_k, \bm{m}_k;\bm{{\lambda}_k}) \log \dfrac{p(\bm{x}_k, \bm{Q}_k, \bm{m}_k, \bm{y}_k|\bm{y}_{1:k-1})}{q(\bm{x}_k, \bm{Q}_k, \bm{m}_k;\bm{\lambda_k})} \text{d}\bm{x}_k \text{d}\bm{Q}_k \text{d}\bm{m}_k}_{\mathcal{B}(\bm{\lambda}_k)}
%\end{split}
%\end{equation}
where the logarithm of the joint PDF $\mathcal{J}_k = \log p(\bm{y}_k, \bm{x}_k, \bm{Q}_k, \bm{m}_k|\bm{y}_{1:k-1})$ can be decomposed as
\begin{equation}
\begin{split}
\mathcal{J}_k =& \log \mathcal{N}(\bm{y}_k|h(\bm{x}_k), \bm{R}_k) + \log \mathcal{N}(\bm{x}_k|\bm{m}_k, \bm{Q}_k) \\
& + \log \mathcal{N}(\bm{m}_k|\bm{\hat m}_{k|k-1}, \bm{\Sigma}_{k|k-1}) \\
& + \log \text{IW}(\bm{Q}_k|{\hat u}_{k|k-1}, \bm{U}_{k|k-1}).
\end{split}
\end{equation}

Then the variational ELBO $\mathcal{B}(\bm{\lambda}_k)$ can be written as Eq.~\eqref{eq::ELBOFinal}.
\begin{figure*} [!tp]
\normalsize
\setcounter{MYtempeqncnt1}{\value{equation}}
\setcounter{equation}{23}
\begin{equation}\label{eq::ELBOFinal}
\begin{split}
\mathcal{B}(\bm{\lambda}_k) =& -\dfrac{1}{2}\mathbb{E}_{q(\bm{x}_k)}\left[(\bm{y}_k - h(\bm{x}_k))^T \bm{R}_k^{-1} (\bm{y}_k - h(\bm{x}_k))\right] \\
                        & -\dfrac{1}{2}\mathbb{E}_{q(\bm{x}_k), q(\bm{m}_k), q(\bm{Q}_k)}\left[(\bm{x}_k - \bm{m}_k)^T \bm{Q}_k^{-1} (\bm{x}_k - \bm{m}_k)\right] \\
                        & -\dfrac{1}{2}\mathbb{E}_{q(\bm{m}_k)}\left[(\bm{m}_k - \bm{\hat m}_{k|k-1})\bm{\Sigma}_{k|k-1}^{-1} (\bm{m}_k - \bm{\hat m}_{k|k-1}) \right] \\
                        & -\dfrac{1}{2}({\hat u}_{k|k-1} + n_x + 2)\mathbb{E}_{q(\bm{Q}_k)}[\log|\bm{Q}_k|] - \dfrac{1}{2}\text{Tr}(\bm{U}_{k|k-1} \mathbb{E}_{q(\bm{Q}_k)}[\bm{Q}_k^{-1}]) \\
                        & - \mathbb{E}_{q(\bm{x}_k)}[\log q(\bm{x}_k)] - \mathbb{E}_{q(\bm{m}_k)}[\log q(\bm{m}_k)] - \mathbb{E}_{q(\bm{Q}_k)}[\log q(\bm{Q}_k)] + \text{const.}
\end{split}
\end{equation}
\hrulefill
\setcounter{equation}{\value{MYtempeqncnt1}}
\end{figure*}
\addtocounter{equation}{1}
Next we derive $q(\bm{x}_k;\bm{\hat x}_{k|k}, \bm{P}_{k|k})$, $q(\bm{Q}_k;\bm{\hat u}_{k|k}, \bm{U}_{k|k})$ and $q(\bm{m}_k;\bm{\hat m}_{k|k}, \bm{\Sigma}_{k|k})$.

\emph{1. Derivations of $q(\bm{x}_k;\bm{\hat x}_{k|k}, \bm{P}_{k|k})$:}  Rewrite the ELBO $\mathcal{B}(\bm{\lambda}_k)$ as the function of state $\bm{x}_k$ and omit the rest terms that are independent of $\bm{x}_k$, denoted by $\mathcal{B}_x(\bm{\hat x}_{k|k}, \bm{P}_{k|k})$. See Eq.~\eqref{eq::ELBOx}.
\begin{figure*} [!tp]
\normalsize
\setcounter{MYtempeqncnt2}{\value{equation}}
\setcounter{equation}{24}
\begin{equation}\label{eq::ELBOx}
\begin{split}
\mathcal{B}_x(\bm{\hat x}_{k|k}, \bm{P}_{k|k}) =& \underbrace{-\dfrac{1}{2}\mathbb{E}_{q(\bm{x}_k)}\left[(\bm{y}_k - h(\bm{x}_k))^T \bm{R}_k^{-1} (\bm{y}_k - h(\bm{x}_k))\right]}_{\mathcal{D}_x(\bm{\hat x}_{k|k}, \bm{P}_{k|k})} \\
                        & \underbrace{-\dfrac{1}{2}\mathbb{E}_{q(\bm{x}_k), q(\bm{m}_k), q(\bm{Q}_k)}\left[(\bm{x}_k - \bm{m}_k)^T \bm{Q}_k^{-1} (\bm{x}_k - \bm{m}_k)\right] -  \mathbb{E}_{q(\bm{x}_k)}[\log q(\bm{x}_k)]}_{\mathcal{E}_x(\bm{\hat x}_{k|k}, \bm{P}_{k|k})}.
\end{split}
\end{equation}
\hrulefill
\setcounter{equation}{\value{MYtempeqncnt2}}
\end{figure*}
\addtocounter{equation}{1}
The ELBO $\mathcal{B}_x(\bm{\hat x}_{k|k}, \bm{P}_{k|k})$ in Eq.~(\ref{eq::ELBOx}) can be divided into two terms, where $\mathcal{D}_x(\bm{\hat x}_{k|k}, \bm{P}_{k|k})$ contains the nonlinear function $h_k(\bm{x}_k)$ that makes the optimization of  natural gradient intractable, and $\mathcal{E}_x(\bm{\hat x}_{k|k}, \bm{P}_{k|k})$ is tractable. To deal with the intractable terms $\mathcal{D}_x(\bm{\hat x}_{k|k}, \bm{P}_{k|k})$, common approaches typically involve making tractable approximations to the nonlinear function $h_k(\bm{x}_k)$ and then compute the expectation. For example, one such approximation would pick a point $\bm{x}_k^*$ and make the first-order Taylor approximation $h_k(\bm{x}_k) \approx h(\bm{x}_k^*) + H(\bm{x}_k^*)(\bm{x}_k - \bm{x}_k^*)$ with $H(\bm{x}_k^*)$ being the Jacobian matrix of $h_k(\bm{x}_k)$. One then replaces $h_k(\bm{x}_k)$ in $\mathcal{D}_x(\bm{\hat x}_{k|k}, \bm{P}_{k|k})$ with the linear approximation and optimizes $\mathcal{B}_x(\bm{\hat x}_{k|k}, \bm{P}_{k|k})$.
As stated in~\cite{2015KhanKL}, this approximation is not tight and can result in a lousy performance. In this case, the resulting update of $q(\bm{x}_k)$ is the same as the iterative EKF with unknown model parameters.

Since the natural gradient of $\mathcal{D}_x(\bm{\hat x}_{k|k}, \bm{P}_{k|k})$ is not available in the
closed-form,
we calculate the gradient $\nabla \mathcal{B}_x(\bm{\hat x}_{k|k}, \bm{P}_{k|k})$ by sampling from the approximate PDF $q(\bm{x}_k)$. That is,
\begin{equation}\label{eq::gradientsS}
  \nabla \mathcal{B}_x(\bm{\hat x}_{k|k}, \bm{P}_{k|k}) = \nabla \mathcal{D}_x(\bm{\hat x}_{k|k}, \bm{P}_{k|k}) +
  \nabla \mathcal{E}_x(\bm{\hat x}_{k|k}, \bm{P}_{k|k}).
\end{equation}

In the following, we calculate $\nabla \mathcal{D}_x(\bm{\hat x}_{k|k}, \bm{P}_{k|k})$ and $\nabla \mathcal{E}_x(\bm{\hat x}_{k|k}, \bm{P}_{k|k})$. Let $S(\bm{x}_k) = -\dfrac{1}{2} (\bm{y}_k - h(\bm{x}_k))^T \bm{R}_k^{-1} (\bm{y}_k - h(\bm{x}_k))$, we have
\begin{equation}\label{eq::gradientsD}
  \nabla \mathcal{D}_x(\bm{\hat x}_{k|k}, \bm{P}_{k|k}) =  \mathbb{E}_{q(\bm{x}_k)} \left[ S(\bm{x}_k) \nabla \log q(\bm{x}_k)\right],
\end{equation}
where the identity $\nabla q(\bm{x}_k) = q(\bm{x}_k) \nabla \log q(\bm{x}_k)$ is used.

The gradient in Eq.~(\ref{eq::gradientsD}) can be approximated by Monte Carlo integration with samples from the variational distribution,
\begin{equation}\label{eq::gradientsD-1}
\begin{split}
& \mathbb{E}_{q(\bm{x}_k)} \left[ S(\bm{x}_k) \nabla \log q(\bm{x}_k)\right] \\
& \approx \dfrac{1}{N} \sum_{i = 1}^N S(\bm{x}_k^i) \nabla \log q(\bm{x}^i_k), \,\, \text{where} \,\, \bm{x}_k^i \overset{iid}{\sim} q(\bm{x}_k),
\end{split}
\end{equation}
where $N$ is the number of samples. The variance reduction methods, which introduce a \emph{control variates} $G(\bm{x}_k)$ that is highly correlated with $S(\bm{x}_k)$ but has an analytic expectation, is employed to reduce the variance of the Monte Carlo integration. Using the control variate $G(\bm{x}_k)$, the gradient in Eq.~(\ref{eq::gradientsD}) is rewritten as
\begin{equation}\label{eq::gradientsD-2}
\begin{split}
& \nabla \mathcal{D}_x(\bm{\hat x}_{k|k}, \bm{P}_{k|k}) \\
& = \mathbb{E}_{q(\bm{x}_k)}\left[(S(\bm{x}_k) - \alpha_k G(\bm{x}_k)) \nabla \log q(\bm{x}_k) \right] \\
& + \alpha_k \nabla \mathbb{E}_{q(\bm{x}_k)}[G(\bm{x}_k)],
\end{split}
\end{equation}
where $\alpha_k = \text{Cov}(S, G) / \text{Var}(G)$ is set to minimize the variance.

In the design of the control variate $G(\bm{x}_k)$, it is required that it is an approximation of $S(\bm{x}_t)$ and its expectation $\mathbb{E}_{q(\bm{x}_k)}[G(\bm{x}_k)]$ is tractable. Based on the EKF framework where the nonlinear function $h(\bm{x}_k)$ is approximated by first-order Taylor expansion at the expectation value $\bm{\hat x}_{k|k}$ of $q(\bm{x}_k)$, i.e., $h(\bm{x}_k) \approx \tilde{h}(\bm{\hat x}_{k|k-1},\bm{x}_k) \triangleq h(\bm{\hat x}_{k|k-1}) + H(\bm{\hat x}_{k|k-1})(\bm{x}_k - \bm{\hat x}_{k|k-1})$, the control variate $G(\bm{x}_t)$ can be designed as
\begin{equation}\label{eq::contral}
  G(\bm{x}_k) = -\dfrac{1}{2}(\bm{y}_k - \tilde{h}(\bm{\hat x}_{k|k-1},\bm{x}_k))^T \bm{R}_k^{-1} (\bm{y}_k - \tilde{h}(\bm{\hat x}_{k|k-1},\bm{x}_k)).
\end{equation}
Substituting Eq.~(\ref{eq::contral}) into Eq.~(\ref{eq::gradientsD-2}), the gradients $\nabla_{\bm{\hat x}_{k|k}} \mathcal{D}_x(\bm{\hat x}_{k|k}, \bm{P}_{k|k})$ and $\nabla_{\bm{P}_{k|k}} \mathcal{D}_x(\bm{\hat x}_{k|k}, \bm{P}_{k|k})$ are given as Eq.~\eqref{eq::gradientFinal}.
\begin{figure*} [!tp]
\normalsize
\setcounter{MYtempeqncnt3}{\value{equation}}
\setcounter{equation}{30}
\begin{equation}\label{eq::gradientFinal}
\begin{split}
  \nabla_{\bm{\hat x}_{k|k}} \mathcal{D}_x(\bm{\hat x}_{k|k}, \bm{P}_{k|k}) =& \dfrac{1}{N} \sum_{i=1}^N \left(S(\bm{x}_k^i) - G(\bm{x}_k^i)\right) \left(\bm{P}_{k|k}^{-1}\bm{x}_k^i - \bm{P}_{k|k}^{-1}\bm{\hat x}_{k|k}\right) \\
  &+ H(\bm{\hat x}_{k|k-1})^T \bm{R}_k^{-1}\left(\bm{y}_k - h(\bm{\hat x}_{k|k-1}) - H(\bm{\hat x}_{k|k-1})(\bm{\hat x}_{k|k} - \bm{\hat x}_{k|k-1})\right),\\
  \nabla_{\bm{P}_{k|k}} \mathcal{D}_x(\bm{\hat x}_{k|k}, \bm{P}_{k|k}) =& \dfrac{1}{2N} \sum_{i=1}^N \left(S(\bm{x}_k^i) - G(\bm{x}_k^i)\right) \left(\bm{P}_{k|k}^{-1}(\bm{x}_k^i - \bm{\hat x}_{k|k})(\bm{x}_k^i - \bm{\hat x}_{k|k})^T\bm{P}_{k|k}^{-1} - \bm{P}_{k|k}^{-1}\right) \\
  &- \dfrac{1}{2} H(\bm{\hat x}_{k|k-1})^T \bm{R}_k^{-1}H(\bm{\hat x}_{k|k-1}).
\end{split}
\end{equation}
\hrulefill
\setcounter{equation}{\value{MYtempeqncnt3}}
\end{figure*}
\addtocounter{equation}{1}
The term $\mathcal{E}_x(\bm{\hat x}_{k|k}, \bm{P}_{k|k})$ in Eq.~(\ref{eq::ELBOx}) can be written as
\begin{equation}\label{eq::gradientE}
\begin{split}
  \mathcal{E}_x(\bm{\hat x}_{k|k}, \bm{P}_{k|k}) = & -\dfrac{1}{2}{\hat u}_{k|k}[(\bm{\hat x}_{k|k} - \bm{\hat m}_{k|k})^T
  \bm{U}_{k|k}^{-1}(\bm{\hat x}_{k|k} - \bm{\hat m}_{k|k})] \\
  & + \dfrac{1}{2} \log |\bm{P}_{k|k}|.
\end{split}
\end{equation}
Then, the gradients $\nabla_{\bm{\hat x}_{k|k}} \mathcal{E}_x(\bm{\hat x}_{k|k}, \bm{P}_{k|k})$ and $\nabla_{\bm{P}_{k|k}} \mathcal{E}_x(\bm{\hat x}_{k|k}, \bm{P}_{k|k})$ are given as
\begin{equation}\label{eq::gradientEFinal}
\begin{split}
  \nabla_{\bm{\hat x}_{k|k}} \mathcal{E}_x(\bm{\hat x}_{k|k}, \bm{P}_{k|k}) =&  -{\hat u}_{k|k}
  \bm{U}_{k|k}^{-1}(\bm{\hat x}_{k|k} - \bm{\hat m}_{k|k}), \\
  \nabla_{\bm{P}_{k|k}} \mathcal{E}_x(\bm{\hat x}_{k|k}, \bm{P}_{k|k}) =&~ \dfrac{1}{2}\bm{P}_{k|k}^{-1} - \dfrac{1}{2}{\hat u}_{k|k} \bm{U}_{k|k}^{-1}.
\end{split}
\end{equation}

Substituting Eq.~(\ref{eq::gradientFinal}) and Eq.~(\ref{eq::gradientEFinal}) into Eq.~(\ref{eq::gradientsS}), 
we have Eq.~\eqref{eq::gradientX}.
\begin{figure*} [!tp]
\normalsize
\setcounter{MYtempeqncnt4}{\value{equation}}
\setcounter{equation}{33}
\begin{equation}\label{eq::gradientX}
\begin{split}
  \nabla_{\bm{\hat x}_{k|k}} \mathcal{B}_x(\bm{\hat x}_{k|k}, \bm{P}_{k|k}) =&  \dfrac{1}{N} \sum_{i=1}^N \left(S(\bm{x}_k^i) - G(\bm{x}_k^i)\right) \left(\bm{P}_{k|k}^{-1}\bm{x}_k^i - \bm{P}_{k|k}^{-1}\bm{\hat x}_{k|k}\right)  - {\hat u}_{k|k} \bm{U}_{k|k}^{-1}(\bm{\hat x}_{k|k} - \bm{\hat m}_{k|k})\\
  &+ H(\bm{\hat x}_{k|k-1})^T \bm{R}_k^{-1}\left(\bm{y}_k - h(\bm{\hat x}_{k|k}) - H(\bm{\hat x}_{k|k-1})(\bm{\hat x}_{k|k} - \bm{\hat x}_{k|k-1})\right), \\
  \nabla_{\bm{P}_{k|k}} \mathcal{B}_x(\bm{\hat x}_{k|k}, \bm{P}_{k|k}) =&~ \dfrac{1}{2N} \sum_{i=1}^N \left(S(\bm{x}_k^i) - G(\bm{x}_k^i)\right) \left(\bm{P}_{k|k}^{-1}(\bm{x}_k^i - \bm{\hat x}_{k|k})(\bm{x}_k^i - \bm{\hat x}_{k|k})^T\bm{P}_{k|k}^{-1} - \bm{P}_{k|k}^{-1}\right) \\
  &- \dfrac{1}{2} H(\bm{\hat x}_{k|k})^T \bm{R}_k^{-1}H(\bm{\hat x}_{k|k}) + \dfrac{1}{2}\bm{P}_{k|k}^{-1} -\dfrac{1}{2}{\hat u}_{k|k}\bm{U}_{k|k}^{-1}.
\end{split}
\end{equation}
\hrulefill
\setcounter{equation}{\value{MYtempeqncnt4}}
\end{figure*}
\addtocounter{equation}{1}

According to Eq.~(\ref{eq::lambda}), the variational parameters $\bm{\hat x}_{k|k}, \bm{P}_{k|k}$ are updated iteratively as
\begin{equation}\label{eq::posteriorX}
  \begin{split}
  \bm{\hat x}_{k|k}^{(n+1)} = \bm{\hat x}_{k|k}^{(n)} + \bm{\rho}_x^{(n)} \nabla_{\bm{\hat x}_{k|k}} \mathcal{B}_x\left(\bm{\hat x}^{(n)}_{k|k}, \bm{P}^{(n)}_{k|k}\right), \\
  \bm{P}_{k|k}^{(n+1)} = \bm{P}_{k|k}^{(n)} + \bm{\rho}_x^{(n)} \nabla_{\bm{P}_{k|k}} \mathcal{B}_x\left(\bm{\hat x}^{(n)}_{k|k}, \bm{P}^{(n)}_{k|k}\right).
  \end{split}
\end{equation}

\emph{2. Derivations of $q(\bm{m}_k; \bm{\hat m}_{k|k}, \bm{\Sigma}_{k|k})$:} Rewrite the ELBO $\mathcal{B}(\bm{\lambda}_k)$ in Eq.~(\ref{eq::ELBOFinal}) as the function of $\bm{m}_k$ and omit the rest terms that are independent of $\bm{m}_k$, denoted by $\mathcal{B}_m(\bm{\hat m}_{k|k}, \bm{\Sigma}_{k|k})$. See Eq.~\eqref{eq::ELBOm}.
\begin{figure*} [!tp]
\normalsize
\setcounter{MYtempeqncnt5}{\value{equation}}
\setcounter{equation}{35}
\begin{equation}\label{eq::ELBOm}
\begin{split}
\mathcal{B}_m(\bm{\hat m}_{k|k}, \bm{\Sigma}_{k|k}) =& \underbrace{-\dfrac{1}{2}\mathbb{E}_{q(\bm{m}_k)}\left[(\bm{m}_k - \bm{\hat m}_{k|k-1})^T \bm{\Sigma}_{k|k-1}^{-1} (\bm{m}_k - \bm{\hat m}_{k|k-1})\right]}_{\mathcal{D}_m(\bm{\hat m}_{k|k}, \bm{\Sigma}_{k|k})} \\
                        & \underbrace{-\dfrac{1}{2}\mathbb{E}_{q(\bm{x}_k), q(\bm{m}_k), q(\bm{Q}_k)}\left[(\bm{x}_k - \bm{m}_k)^T \bm{Q}_k^{-1} (\bm{x}_k - \bm{m}_k)\right] -  \mathbb{E}_{q(\bm{m}_k)}[\log q(\bm{m}_k)]}_{\mathcal{E}_m(\bm{\hat m}_{k|k}, \bm{\Sigma}_{k|k})}.
\end{split}
\end{equation}
\hrulefill
\setcounter{equation}{\value{MYtempeqncnt5}}
\end{figure*}
\addtocounter{equation}{1}
The natural gradient of ELBO $\mathcal{B}_m(\bm{\hat m}_{k|k}, \bm{\Sigma}_{k|k})$ is tractable, which is derived directly as follows. For $\mathcal{D}_m(\bm{\hat m}_{k|k}, \bm{\Sigma}_{k|k})$, its gradients $\nabla_{\bm{\hat m}_{k|k}} \mathcal{D}_m(\bm{\hat m}_{k|k}, \bm{\Sigma}_{k|k})$ and $\nabla_{\bm{\Sigma}_{k|k}} \mathcal{D}_m(\bm{\hat m}_{k|k}, \bm{\Sigma}_{k|k})$ are
\begin{equation}\label{eq::gradientsM}
  \begin{split}
  \nabla_{\bm{\hat m}_{k|k}} \mathcal{D}_m(\bm{\hat m}_{k|k}, \bm{\Sigma}_{k|k}) =& -\bm{\Sigma}_{k|k-1}^{-1}(\bm{\hat m}_{k|k} - \bm{\hat m}_{k|k-1}), \\
  \nabla_{\bm{\Sigma}_{k|k}} \mathcal{D}_m(\bm{\hat m}_{k|k}, \bm{\Sigma}_{k|k}) =& -\dfrac{1}{2}\bm{\Sigma}_{k|k-1}^{-1}.
  \end{split}
\end{equation}
For $\mathcal{E}_m(\bm{\hat m}_{k|k}, \bm{\Sigma}_{k|k})$, its gradients $\nabla_{\bm{\hat m}_{k|k}} \mathcal{E}_m(\bm{\hat m}_{k|k}, \bm{\Sigma}_{k|k})$ and $\nabla_{\bm{\Sigma}_{k|k}} \mathcal{E}_m(\bm{\hat m}_{k|k}, \bm{\Sigma}_{k|k})$ are
\begin{equation}\label{eq::gradientsM}
  \begin{split}
  \nabla_{\bm{\hat m}_{k|k}} \mathcal{E}_m(\bm{\hat m}_{k|k}, \bm{\Sigma}_{k|k}) =& {\hat u}_{k|k}
  \bm{U}_{k|k}^{-1}(\bm{\hat x}_{k|k} - \bm{\hat m}_{k|k}), \\
  \nabla_{\bm{\Sigma}_{k|k}} \mathcal{E}_m(\bm{\hat m}_{k|k}, \bm{\Sigma}_{k|k}) =& \dfrac{1}{2}\bm{\Sigma}_{k|k}^{-1} -\dfrac{1}{2}\hat u_{k|k}\bm{U}_{k|k}^{-1}.
  \end{split}
\end{equation}
Let the gradients of ELBO $\mathcal{B}_m(\bm{\hat m}_{k|k}, \bm{\Sigma}_{k|k})$ equal to zeros.
The optimal variational parameters $\bm{\hat m}_{k|k}, \bm{\Sigma}_{k|k}$ are obtained as
\begin{equation}\label{eq::posteriorM}
  \begin{split}
        \bm{\hat m}_{k|k} &= \bm{\hat m}_{k|k-1} + \bm{G}_k(\bm{\hat x}_{k|k} - \bm{\hat m}_{k|k-1}), \\
        \bm{\Sigma}_{k|k} &= \bm{\Sigma}_{k|k-1} - \bm{G}_k \bm{\Sigma}_{k|k-1}
  \end{split}
\end{equation}
with
\begin{equation}\label{eq::GainM}
  \bm{G}_k = \bm{\Sigma}_{k|k-1} \left(\bm{\Sigma}_{k|k-1} + \bm{U}_{k|k} / \hat u_{k|k} \right)^{-1}.
\end{equation}

\emph{3. Derivations of $q(\bm{Q}_k; \bm{\hat u}_{k|k}, \bm{U}_{k|k})$:} Rewrite the ELBO $\mathcal{B}(\bm{\lambda}_k)$ in Eq.~(\ref{eq::ELBOFinal}) as the function of $\bm{Q}_k$ and omit the rest terms that are independent of $\bm{Q}_k$, denoted by $\mathcal{B}_Q(\hat u_{k|k}, \bm{U}_{k|k})$.
See Eq.~\eqref{eq::ELBOQ}.
\begin{figure*} [!tp]
\normalsize
\setcounter{MYtempeqncnt6}{\value{equation}}
\setcounter{equation}{40}
\begin{equation}\label{eq::ELBOQ}
\begin{split}
\mathcal{B}_Q(\hat u_{k|k}, \bm{U}_{k|k}) =& - \dfrac{1}{2}(\hat u_{k|k-1} + n_k + 2)\mathbb{E}_{q(\bm{Q}_k)}\log |\bm{Q}_k| - \dfrac{1}{2}\text{Tr}(\bm{U}_{k|k-1}\mathbb{E}_{q(\bm{Q}_k)}[\bm{Q}_k^{-1}])] \\
                        & -\dfrac{1}{2}\mathbb{E}_{q(\bm{x}_k), q(\bm{m}_k), q(\bm{Q}_k)}\left[(\bm{x}_k - \bm{m}_k)^T \bm{Q}_k^{-1} (\bm{x}_k - \bm{m}_k)\right] -  \mathbb{E}_{q(\bm{Q}_k)}[\log q(\bm{Q}_k)].
\end{split}
\end{equation}
\hrulefill
\setcounter{equation}{\value{MYtempeqncnt6}}
\end{figure*}
\addtocounter{equation}{1}
Note that the ELBO $\mathcal{B}_Q(\hat u_{k|k}, \bm{U}_{k|k})$ can be rewritten as
\begin{equation}\label{eq::ELBOQNew}
\begin{split}
\mathcal{B}_Q(\hat u_{k|k}, \bm{U}_{k|k}) =& - \dfrac{1}{2}(\hat u_{k|k-1} + 1 - \hat u_{k|k})\mathbb{E}_{q(\bm{Q}_k)}\log |\bm{Q}_k| \\
&- \dfrac{1}{2}\text{Tr}\left(\bm{U}_{k|k-1} \!+ \! \bm{C}_k \!-\! \bm{U}_{k|k} \right)\mathbb{E}_{q(\bm{Q}_k)}[\bm{Q}_k^{-1}]
\end{split}
\end{equation}
with
\begin{equation}\label{eq::C}
  \begin{split}
        \bm{C}_k =& \mathbb{E}_{q(\bm{x}_k), q(\bm{m}_k))}[(\bm{x}_{k} - \bm{m}_{k})(\bm{x}_{k} - \bm{m}_{k})^T] \\
            =& (\bm{\hat x}_{k|k} - \bm{\hat m}_{k|k})(\bm{\hat x}_{k|k} - \bm{\hat m}_{k|k})^T + \bm{P}_{k|k} + \bm{\Sigma}_{k|k}.
  \end{split}
\end{equation}
It is seen that the natural gradient of ELBO $\mathcal{B}_Q(\hat u_{k|k}, \bm{U}_{k|k})$ equals to zero when the optimal variational parameters are
\begin{equation}\label{eq::posteriorQ}
  \begin{split}
        \hat u_{k|k} =& \hat u_{k|k-1} + 1,\\
        \bm{U}_{k|k} =& \bm{U}_{k|k-1} + \bm{C}_k.\\
  \end{split}
\end{equation}

\subsection{Summary}
The proposed nonlinear adaptive Kalman filtering algorithm based on stochastic search variational Bayesian, referred as SSVBAKF, for a single time step is summarized in Algorithm 1, which is a recursive solution of joint nonlinear state estimation and model parameters identification, consisting of closed-loop iterations among state estimation $\bm{x}_k$, auxiliary latent variable $\bm{m}_k$ and process noise covariance $\bm{Q}_k$ at each time.

\begin{algorithm}
\caption{\bf{SSVBAKF:} Stochastic search variational nonlinear adaptive Kalman filtering}
\begin{algorithmic}[1]
\REQUIRE Measurement $\bm{y}_k$, approximated posterior PDFs of last time $q(\bm{x}_{k-1})$, $q(\bm{m}_{k-1})$, $q(\bm{Q}_{k-1})$, sample size $N$, and the maximum number of iterations $I_{\text{max}}$;
\ENSURE Approximated posterior PDFs of current time $q(\bm{x}_{k})$, $q(\bm{m}_{k})$, $q(\bm{Q}_{k})$;
\STATE \underline{\textbf{Time prediction:}}
\STATE Calculate the prior PDF $p(\bm{m}_{k}|\bm{y}_{1:k-1}) = \mathcal{N}(\bm{m}|\bm{\hat m}_{k|k-1}, \bm{\Sigma}_{k|k-1})$ via Eq.~(\ref{eq::predMFinal}).
\STATE Calculate the prior PDF $p(\bm{Q}_k|\bm{y}_{1:k-1}) = \text{IW}(\bm{Q}_k|\hat u_{k|k-1}, \bm{U}_{k|k-1})$ via Eq.~(\ref{eq::predQFinal}).
\STATE \underline{\textbf{Measurement update:}}
\STATE \emph{Initialization:} $q^{(0)}(\bm{x}_k) = p(\bm{x}_k|y_{1:k-1})$, $q^{(0)}(\bm{m}_k) = p(\bm{m}_k|y_{1:k-1})$, $q^{(0)}(\bm{Q}_k) = p(\bm{Q}_k|y_{1:k-1})$
\FOR{each iteration $n = 1 : I_{\text{max}}$  }
\STATE  {\emph{Update posterior $q(\bm{x}_k)$:}} Sample $\bm{x}_k^{i} \overset{\text{iid}}{\sim} q^{(n-1)}(\bm{x}_k)$ for $i = 1, \ldots, N$, compute the gradients $\nabla_{\bm{\hat x}_{k|k}} \mathcal{B}_x(\bm{\hat x}_{k|k}, \bm{P}_{k|k})$ and $\nabla_{\bm{P}_{k|k}} \mathcal{B}_x(\bm{\hat x}_{k|k}, \bm{P}_{k|k})$ as in Eq.~(\ref{eq::gradientX}), and update variational parameters $\bm{\hat x}_{k|k}$ and $\bm{P}_{k|k}$ via Eq.~(\ref{eq::posteriorX}).
\STATE  {\emph{Update posterior $q(\bm{m}_k)$:}} update variational parameters $\bm{\hat m}_{k|k}$ and $\bm{\Sigma}_{k|k}$ via Eq.~(\ref{eq::posteriorM}).
\STATE {\emph{Update posterior $q(\bm{Q}_k)$:}} update variational parameters ${\hat u}_{k|k}$ and $\bm{U}_{k|k}$ via Eq.~(\ref{eq::posteriorQ}).
\ENDFOR
\end{algorithmic}
\label{alg}
\end{algorithm}

\section{Results for Simulated Data}\label{sec:simulation}
In this section, we evaluate the performance of the proposed SSVBAKF algorithm in three different simulated scenarios of radar maneuvering target tracking.
The results for real scenarios will be presented in the next section.

\subsection{Simulation Scenarios}
In all the three scenarios, the target state $\bm{x}_k = [x_k, \dot x_k, y_k, \dot y_k]^T$ consists of target position $[x_k, y_k]^T$ and target velocity $[\dot x_k, \dot y_k]^T$. The nearly-constant velocity model is used to describe the typical target trajectories. The transition matrix $\bm{F}_k$~(in this case $f_k(\cdot)$ is linear) and the process noise covariance $\bm{Q}_k$ are given by
\begin{equation*}
  \bm{F}_k = \bm{I}_{2} \otimes \begin{bmatrix}
     1 & T \\
     0 & 1 \\
   \end{bmatrix}, \qquad
   \bm{Q}_k = \delta_k^2 \bm{I}_2 \otimes \begin{bmatrix}
T^3/3 & T^2/2 \\
T^2/2 & T \\
\end{bmatrix}
\end{equation*}
where $\bm{I}_2$ is $2\times 2$ identity matrix, $\delta_k^2$ is process noise level, and $T$ is the sampling period.

The target is detected by a 2D radar~(located at the origin $[0m, 0m]$) which provides measurements $\bm{y}_k = [r_k, \theta_k]^T$ consisting of range $r_k$ and azimuth $\theta_k$.
The nonlinear measurement function $\bm{h}_k(\cdot)$ is given by
\begin{equation*}
  r_k = \sqrt{x_k^2 + y_k^2}, \qquad \theta_k = \arctan\left(y_k/x_k\right).
\end{equation*}

Three different maneuvering target tracking scenarios are considered as follows:
\begin{itemize}
  \item \textbf{S1}: \emph{Aerial target tracking with air surveillance radar}.
  \item \textbf{S2}: \emph{Vehicle target tracking with ground surveillance radar}.
  \item \textbf{S3}: \emph{Ship target tracking with marine surveillance radar}.
\end{itemize}

Given the initial target state $\bm{x}_0$, the ground truth of the simulated target trajectory is generated by sampling from the Gaussian distribution recursively, i.e., $\bm{x}_k \sim N(\bm{F}_k\bm{x}_{k-1}, \bm{Q}_k)$. The corresponding scenario parameters, including the initial state $\bm{x}_0$, process noise level $\delta^2$, measurement noise covariance $\bm{R}$ and sampling period $T$ are given in TABLE \ref{ScenarioPar}. The total simulated steps is 300 whereas the target maneuvers in the middle steps.
More specifically, the true process noise level $\delta_k^2 = \delta^2, k \in [1, 100)$, $\delta_k^2 = 10 \times \delta^2, k \in [100, 199)$ and $\delta_k^2 = \delta^2, k \in [200, 300]$.
\begin{table*}[htp]
    \centering
    \caption{Scenario parameters}
    \label{ScenarioPar}
    \begin{tabular}{c|c|c|c|c}
    \hline \hline
        \textbf{Scenario} & $\bm{x}_0$ & $\delta^2$ & $\bm{R}$ & $T$ \\ \hline
        \textbf{S1} & $[5e^5 m, -100 m/s, 5e^5 m, -100 m/s]^T$ & $10 m^2/s^3$ & $\text{diag}(1e^4 m^2, 5e^{-5} rad^2)$ & 10s \\
        \textbf{S2} & $[1e^4 m, -10 m/s, 1e^4 m, -10 m/s]^T$ & $2.5 m^2/s^3$ & $\text{diag}(1e^2 m^2, 5e^{-6} rad^2)$ & 2s \\
        \textbf{S3} & $[5e^3 m, -5 m/s, 5e^3 m, -5 m/s]^T$ & $1 m^2/s^3$ & $\text{diag}(25 m^2, ~1e^{-6} rad^2)$ & 3s \\
    \hline
    \end{tabular}
\end{table*}

The compared methods, including model-based adaptive filters such as the Interacting Multiple Models (IMM), and noise-based adaptive filters such as the Variational Bayesian Adaptive Cubature Information Filter (VBACIF), are described as follows.
\begin{itemize}
\item IMM~\cite{2005RongIMM}: This is a model-adaptive filter that uses an individual mode for each noise level $\delta_{i}^2 \in {0.1\delta^2, 0.5\delta^2, 1\delta^2, 5\delta^2, 10\delta^2}$ (five models in total).
The mode transition probabilities from mode $i$ to mode $j$ are $P_{ij} = 0.8$ for $i=j$ and $P_{ij}=0.05$ for $i \neq j$. EKF performs the nonlinear measurement update step.
\item VBACIF~\cite{2017DongVariational}: This is a measurement noise covariance adaptive nonlinear filter that models the inverse of measurement noise covariance as a Wishart distribution and approximates the integration of recursive Bayesian estimation by a cubature integration rule. The initial parameters are $v_0 = 1$ and $V_0 = 2 \bm{R}^{-1}$. The initial noise level for all scenarios is $\delta_0^2 = 20m^2/s^3$, the variance decreasing factor is $\beta = 1 - e^{-4}$, and the maximum number of iterations is $I_{\text{max}} = 500$.
\item SSVBAKF: This is a process noise covariance adaptive nonlinear filter that approximates the integration of recursive Bayesian estimation by stochastic search variational Bayes. The initial parameter is $\tau = 3$, the variance decreasing factor is $\beta = 1 - 1e^{-4}$, the number of samples is $N = 1000$, and the maximum number of iterations is $I_{\text{max}} = 500$. The initial noise level for all the three scenarios is $\delta_0^2 = 20 m^2/s^3$.
\end{itemize}

\subsection{Results}
\begin{figure}[htp]
    \centering
    \subfloat[S1]{\includegraphics[width=0.75\linewidth]{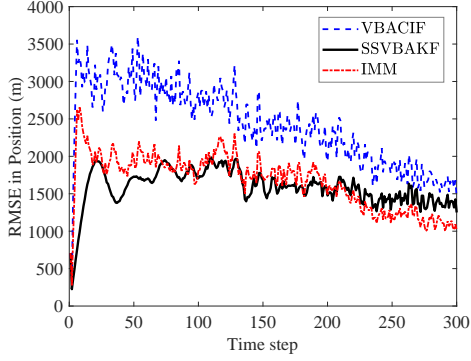}} \\
    \subfloat[S2]{\includegraphics[width=0.75\linewidth]{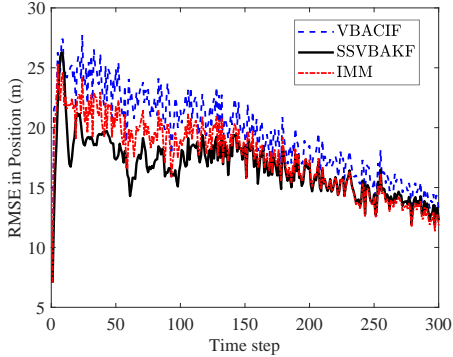}} \\
    \subfloat[S3]{\includegraphics[width=0.75\linewidth]{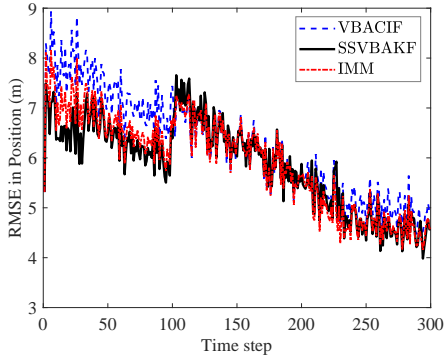}}
    \caption{The curve of RMSE comparison for simulated scenarios}
    \label{Sim target rmse}
\end{figure}

\begin{figure}[htp]
    \centering
    \subfloat[S1]{
        \label{skyelbo}
        \includegraphics[width=0.75\linewidth]{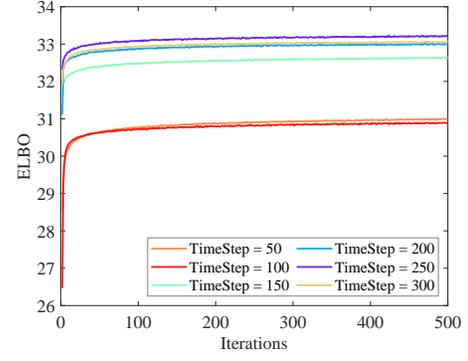}
        } \\
    \subfloat[S2]{
        \label{groundelbo}
        \includegraphics[width=0.75\linewidth]{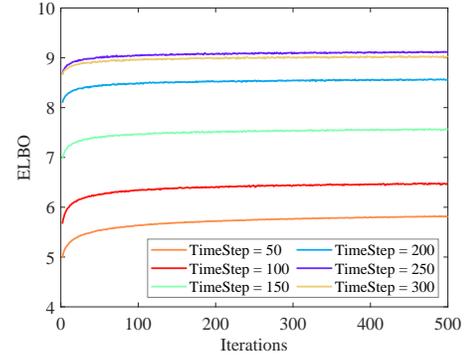}
        }\\
     \subfloat[S3]{
        \label{seaelbo}
        \includegraphics[width=0.75\linewidth]{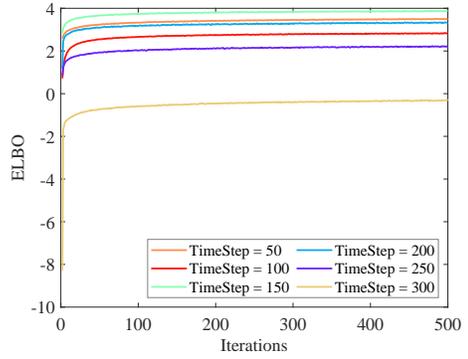}
        }
    \caption{The curves of ELBO for simulated scenarios}
    \label{Sim target ELBO}
\end{figure}

For each scenario, 100 Monte Carlo runs are carried out. The root mean square error~(RMSE) is compared in Fig.~\ref{Sim target rmse}.
It is shown that SSVBAKF has comparable estimation performance with IMM and outperforms VBACIF in all scenarios.
This is because SSVBAKF directly optimizes the joint state estimation and unknown process noise covariance.
In contrast, VBACIF performs the adaptive filtering by dynamically turning the measurement noise covariances, which is indirect.
Meanwhile, IMM has shown satisfactory estimation results under the assumption of a complete set of models, which are not trivial in practice.
The average RMSE over time is shown in Table~\ref{mean sim rmse}, demonstrating that SSVBAKF achieves the best estimate accuracy in all scenarios.
The ELBO curves of SSVBAKF for different iterations are shown in Fig.~\ref{Sim target ELBO}. At each iterative cycle, the ELBO increases with the number of iterations increase, indicating that the iteration procedure in SSVBAKF converges.

\begin{table}[htp]%
    \centering
    \caption{Average RMSE in simulated scenarios}
    \label{mean sim rmse}
    \begin{tabular}{c|ccc}    \hline \hline
        \textbf{Scenario} & \textbf{VBACIF} & \textbf{IMM} & \textbf{SSVBAKF} \\ \hline
        \textbf{S1} & 2427.62 & 1637.30 & $\bm{1584.31}$ \\
        \textbf{S2} & 19.59 & 17.58 & $\bm{16.65}$\\
        \textbf{S3} & 6.26 & 5.94 & $\bm{5.87}$\\ \hline
    \end{tabular}
\end{table}

\section{Results for Real Data}\label{sec:real}
Next, we apply SSVBAKF to real scenarios, including aerial target tracking, vehicle target tracking, and ship target tracking with different types of 2D surveillance radar. GPS provides the ground truth of the target trajectory.

\subsection{Real Scenarios}
\begin{figure}[htp]
    \centering
    \subfloat[R1]{
        \label{realS1}
        \includegraphics[width=0.75\linewidth]{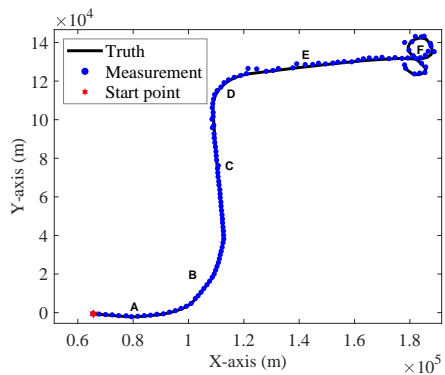}
        } \\
    \subfloat[R2]{
        \label{realS2}
        \includegraphics[width=0.75\linewidth]{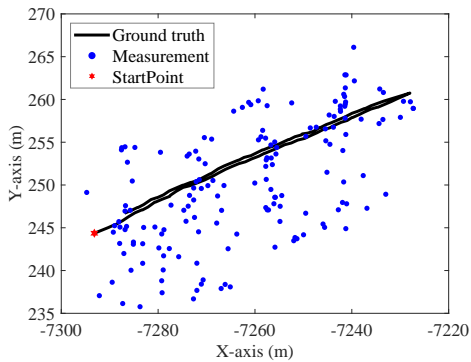}
        } \\
    \subfloat[R3]{
        \label{realS3}
        \includegraphics[width=0.75\linewidth]{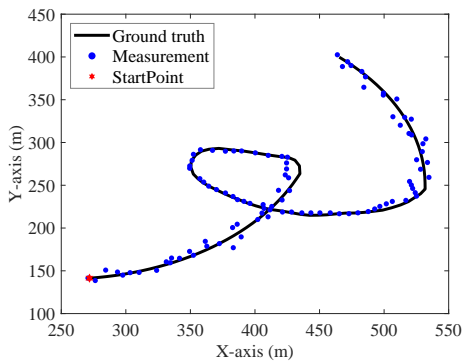}
        }
    \caption{Target trajectories and radar detection for real scenarios}
    \label{all real targets}
\end{figure}

The target trajectory and its corresponding radar detection characteristics for different real scenarios are given as follows.
\begin{itemize}
\item \textbf{R1}: \emph{Aerial target tracking with an air surveillance radar}. The trajectory of the aerial target is shown in Fig.~\ref{realS1}, which can be divided into six segments~(maneuver modes): CV~(segment A), left CT~(segment B), CV~(segment C), right CT~(segment D), CV~(segment E) and figure-eight flight pattern~(segment F). The target is detected by an air surveillance radar. The scanning period $T = 10s$, the detection precision of range and azimuth is 100m and 3.5e-3rad.
\item \textbf{R2}: \emph{Vehicle target tracking with a ground surveillance radar}. The trajectory of the vehicle target is shown in Fig.~\ref{realS2}, which makes a round-trip maneuver. The target is detected by a ground surveillance radar. The scanning period $T = 2s$, the detection precision of range and azimuth is 10m and 2.5e-3rad.
\item \textbf{R3}: \emph{Ship target tracking with a marine surveillance radar}. The trajectory of the ship target is shown in Fig.~\ref{realS3}, which makes turning maneuvers with time-varying turning rates. The target is detected a by marine surveillance radar. The scanning period $T=3s$, the detection precision of range and azimuth is 5m and 1e-3rad.
\end{itemize}

%\begin{figure}[htbp]
%    \centering
%    \subfloat[S1:Airplane target]{
%        \label{realS1}
%        \includegraphics[scale=0.5]{Figure/fig4.eps}
%        }\\
%    \subfloat[S2:Vehicle target]{
%        \label{realS2}
%        \includegraphics[scale=0.5]{Figure/fig5.eps}
%        }\\
%    \subfloat[S3:Ship target]{
%        \label{realS3}
%        \includegraphics[scale=0.5]{Figure/fig6.eps}
%        }
%    \caption{Target trajectories in three real scenarios}
%    \label{all real targets}
% \end{figure}

The parameter settings for all algorithms are the same as in simulated scenarios.

\subsection{Results}
\begin{figure}[htp]
\centering
\subfloat[R1]{
\label{rmses1}
\includegraphics[width=0.75\linewidth]{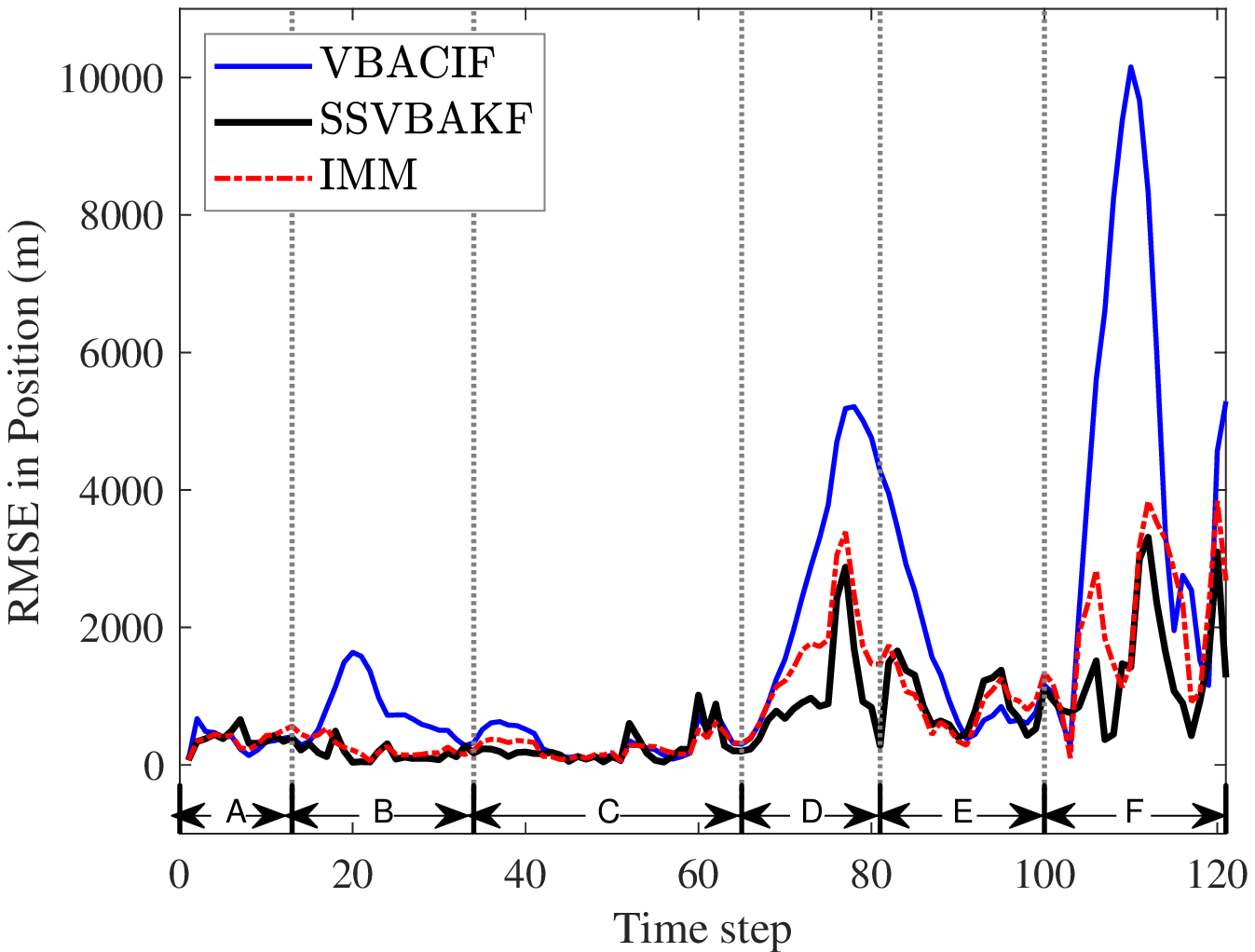}
} \\
\subfloat[R2]{
\label{rmses2}
\includegraphics[width=0.75\linewidth]{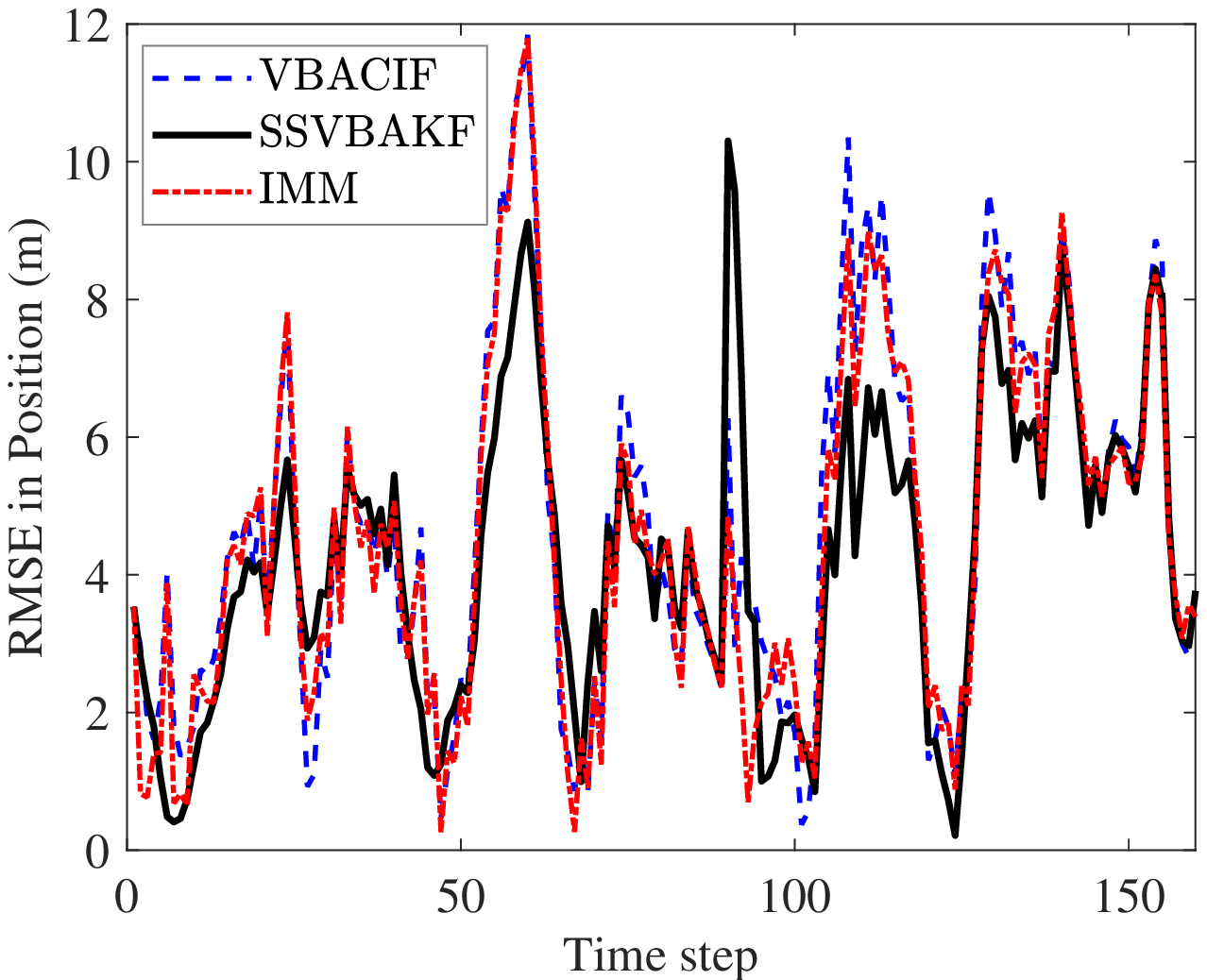}
} \\
\subfloat[R3]{
\label{rmses3}
\includegraphics[width=0.75\linewidth]{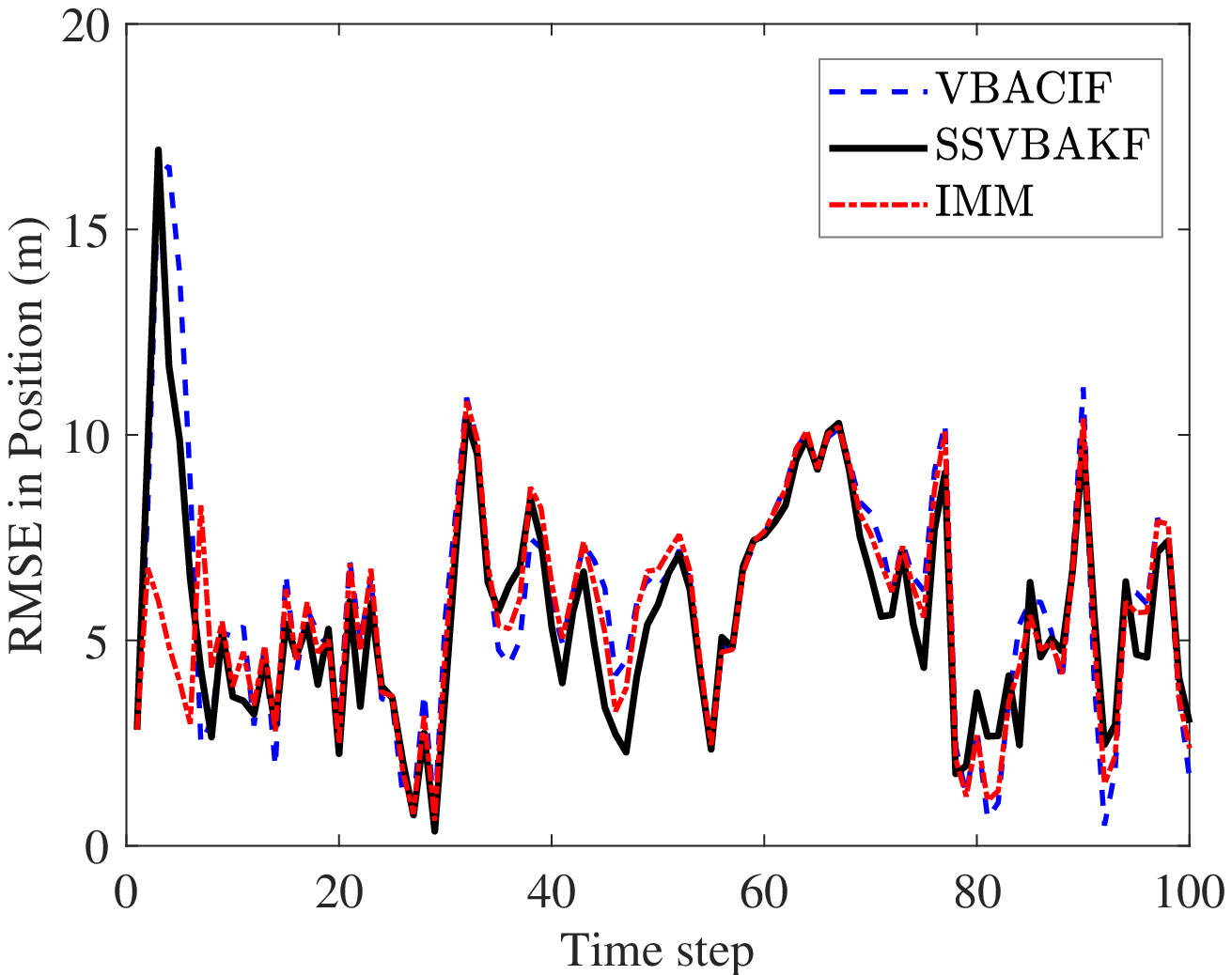}
}
\caption{The curve of RMSE comparison for real scenarios}
\label{real scenarios rmse}
\end{figure}

The RMSE and averaged RMSE over time for real scenarios are shown in Fig.~\ref{real scenarios rmse} and Table~\ref{tab4}, respectively. It is shown that SSVBAKF outperforms IMM and VBACIF in all scenarios.
In Fig.~\ref{rmses1}, we conduct a segment analysis of the RMSE for Scenario R1. The result shows little difference in the RMSE curves among the algorithms in the non-maneuvering and weakly maneuvering segments. However, SSVBAKF has a more obvious advantage in the maneuvering segments. Table \ref{Air Scenario Results} records the RMSE of each algorithm for the air target in different motion modes. The results show that all the three algorithms perform well on filtering  when the target moves in a straight line at approximately uniform speed or when the turning maneuver is small (segments A, B, and C). When the turning maneuver is significant (segments D and F) or the target is accelerated~(segment E), the filtering performance of all the three algorithms decreases. However, SSVBAKF achieves better performance than IMM and VBACIF.
The filtering performance of IMM and VBACIF is more dependent on the motion scenario.
When the target is in CV motion or weak maneuvering (e.g., segments A, B, and C), the RMSE of the VBACIF algorithm is overall lower than that of IMM.
At the same time, IMM performs better when the target is in accelerate motion or high maneuvering (e.g., segments D, E, and F).

Figs. \ref{rmses2} and \ref{rmses3} present the RMSE curves for Scenario R2 and Scenario R3, respectively.
Unlike the aerial target scenario~(R1),
the targets in R2 and R3 move in a more homogeneous pattern (straight lines or turns) and move more slowly (especially the ship target in R3).
Therefore, all the three filtering algorithms can achieve good filtering results.
The RMSEs on filtering are given in Table \ref{tab4}.
%Table \ref{tab4} records the average RMSE values of each algorithm in three scenarios.
The results show that SSVBAKF has achieved the best filtering results in all scenarios.
When the model set of IMM cannot well capture the motion mode of the target,
the filtering accuracy will be deteriorated.
In contrast, SSVBAKF adaptively estimates the process noise covariance without relying on the initial value of $\delta$ or the model set of $\delta$. This adaptive estimation is designed to ensure that the filtering gain stays within a reasonable range by changing the process noise covariance when the target undergoes maneuvering.

\begin{table}[htp]
    \centering
    \caption{Air Scenario Results}
    \label{Air Scenario Results}
    \begin{tabular}{cc|ccc}
    \hline \hline
        \multicolumn{2}{c|}{Target Movement Mode} & \multicolumn{3}{c}{Average RMSE} \\ \hline
    {Segment} & {Time Interval} & {VBACIF} & {IMM} & {SSVBAKF}\\  \hline
        A & $[0, 13)$  & 240.38 & 316.23 & $\bm{230.00}$\\
        B & $[13, 34)$ & 304.35 & 330.69 & $\bm{193.36}$ \\
        C & $[34, 65)$ & 408.14 & 328.01 & $\bm{217.36}$ \\
        D & $[65, 81)$ & 2329.40 & 929.37 & $\bm{255.82}$ \\
        E & $[81, 100)$ & 2553.77 & 1539.92 & $\bm{1297.32}$ \\
        F & $[100, 121)$ & 3162.76 & 1928.70 & $\bm{1108.21}$ \\
    \hline
    \end{tabular}
\end{table}

\begin{table}[htp]%
    \centering
    \caption{Comparison of mean RMSE in real scenarios}
    \label{tab4}
    \begin{tabular}{c|ccc}
    \hline \hline
     \textbf{Scenario} & \textbf{VBACIF} & \textbf{IMM} & \textbf{SSVBAKF} \\ \hline
        \textbf{R1} & 1654.43  & 876.04 & $\bm{651.00}$ \\
        \textbf{R2} & 5.99  & 5.65 & $\bm{5.62}$ \\
        \textbf{R3} & 4.67  & 4.53 & $\bm{4.24}$\\
    \hline
    \end{tabular}
\end{table}

\section{Conclusion}\label{sec:conclusion}
We have proposed an optimisation-based nonlinear adaptive Kalman filtering approach to the problem of maneuvering target tracking, by encapsulating the nonlinear state estimation with uncertain process noise covariance into a variational inference problem. To deal with the problems of nonlinearity and the non-conjugate prior encountered by the mean-filed variational inference, the stochastic search variational inference with an auxiliary latent variable is adopted to offer a flexible, accurate and effective approximation of the joint posterior distribution. The proposed method iteratively estimates the process noise covariance and the target state. The simulated and real experiments demonstrated that the proposed method achieved high accuracy and outstanding robustness in different scenarios.

\bibliographystyle{IEEEtran} %the bibliographystyle
\bibliography{IEEEabrv,Reference} %the references
\end{document}